\documentclass[12pt]{article}
\usepackage{graphics}
\usepackage{natbib}
\usepackage{url}
\usepackage{color}
\usepackage{epstopdf}
\usepackage{amsmath}
\usepackage{amssymb}
\usepackage{booktabs}
\usepackage{graphicx}
\usepackage{xcolor,colortbl}
\usepackage{lipsum}
\usepackage{setspace}
\usepackage{natbib}
\usepackage{multirow}
\usepackage{bm}
\usepackage{environ}
\usepackage{graphicx}
\usepackage{float}
\usepackage{textcomp}
\usepackage{comment}
\usepackage[pagewise]{lineno}

\renewcommand{\baselinestretch}{1.7}

\NewEnviron{NORMAL}{%
    \scalebox{2}{$\BODY$} 
} 

\NewEnviron{SMALL}{%
    \scalebox{0.5}{$\BODY$} 
} 

\begin{document}

\renewcommand{\baselinestretch}{1.5}
\title{Missing data imputation for a multivariate outcome of mixed variable types}

\author{
\small \bf Tuo Wang\\ \small Department of Biostatistics and Medical Informatics, University of Wisconsin-Madison \\ \small Madison, WI 53706, USA \\
\bf \small Rachel Zilinskas \\ \small Statistics and Data Corporation  \\ \small Tempe, AZ 85288, USA \\
\bf \small Ying Li\\ \small Department of Statistics, Data and Analytics, Eli Lilly and Company \\ \small Indianapolis, IN 46285, USA \\
\small \bf Yongming Qu*\\ \small Department of Statistics, Data and Analytics, Eli Lilly and Company \\ \small  Indianapolis, IN 46285, USA\\
\\\\} 


\date{\small \emph{\today}}

\maketitle
\noindent 
{\small *Correspondence: Yongming Qu, Email: qu\_yongming@lilly.com, Department of Global Statistical Sciences,  Eli Lilly and Company, Lilly Corporate Center, Indianapolis, IN 46285, USA. }

\newpage
\noindent\\
{\small {SUMMARY}  
Data collected in clinical trials are often composed of multiple types of variables. For example, laboratory measurements and vital signs are longitudinal data of continuous or categorical variables, adverse events may be recurrent events, and death is a time-to-event variable. Missing data due to patients' discontinuation from the study or as a result of handling intercurrent events using a hypothetical strategy almost always occur during any clinical trial. Imputing these data with mixed types of variables simultaneously is a challenge that has not been studied extensively. In this article, we propose using an approximate fully conditional specification to impute the missing data. Simulation shows the proposed method provides satisfactory results under the assumption of missing at random. Finally, real data from a clinical trial evaluating treatments for diabetes are analyzed to illustrate the potential benefit of the proposed method.
}


\newpage 

\section{Introduction}
\label{sec:intro}

Often, many important variables (endpoints) are collected in clinical trials. In spite of best efforts, missing values cannot be fully prevented, so appropriately handling of missing values is essential for valid statistical inference. For longitudinal outcomes with ignorable missingness, the likelihood-based approach can be used to draw inference. {For multivariate outcomes, the joint modeling approach can be used \citep{rizopoulos2012joint}. Generally, EM algorithm, which requires customized programming, can be used to estimate the parameters in the joint models. Multiple imputation \citep{rubin1987multiple}, a popular method to handle missing values as the estimation procedure for complete data can be used directly after imputation, offers an alternative approach for this problem.} In addition, multiple imputation is a convenient approach for the implementation of pattern mixture models \citep{little1993pattern}. Most imputation packages either only impute missing data for a specific type of variables or for a limited combination of various types of variables. For example, PROC MI in SAS and mice package in R cannot be used to directly impute missing data for time-to-event (TTE) outcomes. 

\cite{zhao2014multiple, zhao2016sensitivity} used the Kaplan-Meier estimator \citep{kaplan1958nonparametric} for the survival function to impute the censored TTE data. This approach assumes independent censoring and cannot incorporate covariates. \cite{lipkovich2016sensitivity} proposed using a piecewise exponential survival function to impute TTE data, with only incorporating baseline covariates. \cite{murad2020imputing} proposed imputing the TTE variable and other continuous variables simultaneously with the continuous variables imputed by fully conditional specification (FCS); however, this method does not allow the continuous variables to depend on the TTE variable in FCS. This is sub-optimal since, for example, a patient with an event of heart failure may indicate a poor 6-minute walking distance (a continuous variable). In addition, \cite{murad2020imputing} did not conduct simulations to evaluate the performance of their imputation method.

Research on imputation for time to recurrent events (TTREs) is sparse. \cite{keene2014missing} proposed using negative binomial regression to impute the number of recurrent events without incorporating postbaseline variables. \cite{tang2018algorithms} proposed several methods to impute TTREs without including covariates. 

In this article, we provide a general framework for imputing missing data for longitudinal multivariate mixture types of continuous, binary, ordinal, TTE, and TTREs using an approximate FCS. {The proposed methods can serve as a fundamental framework to handle the missing data in various clinical trials. This article is arranged as follows. In Section \ref{sec:methods} we lay out the imputation methods. In Section \ref{sec:simulation}, we conduct simulation to assess the performance of the proposed method. As an illustration, a major diabetes clinical trial is analyzed in Section \ref{sec:application}. Section \ref{sec:summary} concludes the paper with summaries and practical considerations in the application of the proposed method.}

\section{Methods} \label{sec:methods}

We consider $K$ outcome variables of mixed types, denoted by $Y_1, Y_2, \ldots, Y_K$. Let $X$ denote a vector of baseline covariates. For simplicity, we do not include the subject index in the variable names. We assume monotone missingness for all variables 
{[Does the missingness for the longitudinal covariate need to be monotone for all variables? For example could a subject have a missed visit at time k that we impute using k-1 and previous time points but use observed values for k+1 and beyond?]} 
in which the missingness is from administrative censoring or patients' discontinuation from the clinical study. For convenience of discussion, we consider the situation of 1 TTE variable ($Y_1$), 1 TTRE variable (${Y}_2$), with the remainder being {binary or normally distributed continuous variables} ($Y_3$, $Y_4$, $\ldots$, $Y_K$). {Handling ordinal and categorical variables will be discussed later.}

Let $T_c$ be the last time the patient is in the study ($T_c$ is either the terminal event time (death) or the censoring time). Assume $Y_1$ is not a terminal event. Let $T_1^*$ be the event time for $Y_1$, which may not be observed in the presence of censoring. Then, the TTE variable can be described as $Y_1=(T_1, I(T_1^*<T_c))$, in which the first component $T_1 = \min(T_1^*, T_c)$ is the time to event or censoring and the second component is the event indicator. Let ${Y}_2 = (T_{2}^{(1)}, T_{2}^{(2)}, \ldots)$ denote the time to recurrent events up to the censoring time $T_c$. 
Let ${Y}_k = (Y_{k1}, \ldots, Y_{kJ})$ denote $J$ repeated post-randomization measures for the $k$-th ($k \ge 3$) variable. 

To incorporate the correlation between different types of variables, we consider using a FCS originally proposed by \cite{van2006fully} for imputation with the conditional distributions as outlined below: 
\begin{enumerate}
    \item The TTE variable will be imputed using a conditional piecewise hazard model. Basically, we assume the survival function in the current time interval is an exponential distribution when conditioning on previous data. 
    \item The missing data for the recurrent event variable ${Y}_2$ are imputed using a Poisson process with a piecewise rate model conditional on data before the current time interval, similar to the conditional piecewise hazard model. 
    \item For a continuous variable, the missing data are imputed using a conditional normal distribution.  
    \item For the binary variable, a conditional logistic regression model can be used for imputation.
\end{enumerate}

We divide the duration of studies into time intervals. A natural partition is to use the time points when the longitudinal data are measured. Let $0=t_0 < t_1 < \ldots < t_J = t_{\max}$ be the time points to partition the study duration and $\Delta t_j = t_{j}-t_{j-1}$ be the duration of $j$-th interval $(t_{j-1}, t_j]$, for $j = 1, \ldots, J$. 

Let ${Z}^{(j)} = \left(Z_1^{(j)}, Z_2^{(j)}, \ldots, Z_K^{(j)}\right)'$, where 
\[
{Z}_1^{(j)} = \left\{ \begin{array}{cl} \left\{\min(T_1-t_{j-1}, t_{j}-t_{j-1}), I(T_1>t_{j})\right\}, & \mbox{if } T_1>t_{j-1} \\ \mbox{Not defined}, & \mbox{otherwise} \end{array} \right.
\]
denotes the time to event/censoring and the censoring status relative to time $t_{j-1}$ conditional on the event not occurring before $t_{j-1}$, $Z_2^{(j)}$, a set of recurrent events in the $j$-th time interval relative to time $t_{j-1}$, $Z_k^{(j)}, k=3,4,\ldots,K$, the value of $Y_k$ at time point $j$ at
the end of the $j$-th time interval. Let $Z_1^{(j*)}$ denote whether a patient experiences an event (for variable $Y_1$) up to time $t_j$, and $Z_2^{(j*)}$ be the number of recurrent events (for variable $Y_2$) in the $j$-th interval. 

Since the missingness is assumed to be monotone, we can impute the missing data in each interval sequentially.  Given the data are observed or imputed up to time point $t_{j-1}$, the missing data in $Z^{(j)}$ can be imputed through a conditional distribution 
\begin{equation} \label{eq:imputation_model}
f\left(Z_{k}^{(j)}|X, Z^{(1*)}, \ldots, Z^{(j-1,*)}; \theta_k^{(j)}\right), \quad k=1,2,\ldots, K; j=1,2,\ldots, J,
\end{equation}
where $Z^{(j*)} = \left(Z_1^{(j*)}, Z_2^{(j*)}, Z_3^{(j)}, \ldots, Z_K^{(j)}\right)'$. 
For simplicity, we do not differentiate the notations between observed and imputed values, and by default, when the data are not observed, the variables automatically denote the imputed values. In addition, we assume the variables $Z_1^{(k)}$, $Z_2^{(k)}$, $\ldots$, $Z_J^{(k)}$ are independent conditional on the data up to time $t_{k-1}$. This assumption is not overly restrictive, as unconditionally these variables are still dependent. 

We assume a constant hazard rate for $Z_{1}^{(j)}$ in the $j$-th interval, so the conditional distribution for $Z_{1}^{(k)}$ is an exponential distribution. For recurrent events $Z_{2}^{(j)}$, we also assume a constant event rate which can be estimated using a gamma frailty model or a negative binomial regression. 

The imputation is an iterative process from time $t_1$ to $t_J$.  Assume the data up to time $t_{j-1}$ are non-missing or imputed in the previous iteration. Then, the parameter $\theta_k^{(j)}$ in Model (\ref{eq:imputation_model}) can be estimated using subjects with non-missing data up to $t_{j}$, denoted by $\hat\theta_k^{(j)}$.
Next, generate a random number $\tilde{\theta}_k^{(j)}$ based on the estimated parameter with consideration of the variability in the estimator $\hat\theta_k^{(j)}$. Most commonly $\tilde{\theta}_k^{(j)} \sim \mathcal{N} \left(\hat{\theta}_k^{(j)}, V_i^{(j)}\right)$, where $V_i^{(j)} = \widehat{\text{Var}}\left(\hat{\theta}_k^{(j)}\right)$. Then, the missing data for $Z_k^{(j)}$ can be imputed by generating a random number from $f\left(Z_{k}^{(j)}|X, Z^{(1*)}, \ldots, Z^{(j-1,*)}; \tilde \theta_k^{(j)}\right)$. More specifically, the imputation for each type of variables is described below:
\begin{enumerate} 
    \item If $Y_k$ ($k\ge 3$) is a continuous variable, the conditional distribution $f\left(Z_{k}^{(j)}|X, Z^{(1*)}, \ldots,\right.$ $\left. Z^{(j-1,*)}; \theta_k^{(j)}\right)$ for the variable $Z_k^{(j)}$ ($Y_k$ at time $t_j$) can be modeled through a multiple linear regression model:
    \[
    Z_{k}^{(j)} = \left\{W^{(j-1)}\right\}' \beta_k^{(j)} + e_k^{(j)}, e_k^{(j)} \sim \mathcal{N}(0, \sigma_k^{2(j)}),
    \]
    where $W^{(j-1)} = \left(1, X', Z^{(1*)'}, Z^{(2*)'}, \ldots, Z^{(j-1,*)'}\right)'$. Let $\theta_k^{(j)} = \left(\beta_k^{(j)}, \sigma_k^{2(j)}\right)$, and $\hat \theta_k^{(j)} = \left(\hat \beta_k^{(j)}, \hat \sigma_k^{2(j)}\right)$ be the estimator for $\theta_k^{(j)}$ using data for patients with $T_c>t_j$. Generate a random vector $\tilde \beta_k^{(j)} \sim \mathcal{N}\left(\hat \beta_k^{(j)}, \tilde \sigma_k^{2(j)}/\hat \sigma_k^{2(j)} \cdot \widehat{\text{Var}}(\hat \beta_k^{(j)})\right)$, where $\tilde \sigma_k^{2(j)} = \hat \sigma_k^{2(j)} (n_k^{(j)}-m_k^{(j)}-1)/\xi^{(j)}$,  $n_k^{(j)}$ is the number of observations included in the regression, $m_k^{(j)}=\mbox{dim}(W^{(j-1)})$ is the number of covariates in the regression model, and $\xi_k^{(j)}$ is a random number generated from $\chi_{n_k^{(j)}-m_k^{(j)}-1}^2$. Then, the missing value is imputed by:
    \[
    \hat Z_k^{(j)} = \{W^{(j-1)}\}' \tilde \beta_k^{(j)} + \tilde e_k^{(j)}, 
    \]
    where $\tilde e_k^{(j)} \sim \mathcal{N}(0, \tilde \sigma_k^{2(j)})$. Note: $\tilde \beta_k^{(j)}$ and $\xi_k^{(j)}$ are only generated once for each $k$ and $j$ for all patients while $\tilde e_k^{(j)}$ is generated independently for each subject. This process was originally outlined by \cite{van2006fully} and described at greater length in the technical details for PROC MI in SAS\textsuperscript{\tiny\textregistered} User Manual.
    \item If $Y_k$ ($k\ge 3$) is a binary variable, the conditional distribution $f(Z_{k}^{(j)}|X, Z^{(1*)}, \ldots,$ $Z^{(j-1,*)}; \theta_k^{(j)})$ for the variable $Z_k^{(j)}$ (the value of $Y_k$ at time $t_j$) can be modeled through a logistic regression model: 
     \[
    \mbox{logit}\left\{E(Z_{k}^{(j)})\right\} = \{W^{(j-1)}\}' \theta_k^{(j)},
    \]
    where $\mbox{logit}(x)=\log(x/(1-x))$. 
    Let $\hat\theta_k^{(j)}$ be the estimator for $\theta_k^{(j)}$ with variance estimator $\widehat{\text{Var}}(\hat\theta_k^{(j)})$ using data for patients with $T_c>t_j$. Generate a random number for the parameter from $\tilde{\theta}_k^{(j)} \sim \mathcal{N}(\hat\theta_k^{(j)}, \widehat{\text{Var}}(\hat\theta_k^{(j)}))$ to account for the variability in the estimator $\hat \theta_k^{(j)}$. Then, the missing value is imputed by a random variable generated from a Bernoulli distribution with mean
    \[
    \hat p_k^{(j)} = \frac{\exp\left(\{W^{(j-1)}\}' \tilde \theta_k^{(j)}\right)}{1 + \exp\left(\{W^{(j-1)}\}' \tilde \theta_k^{(j)}\right)}.
    \]


    \item For the imputation of the TTE variable $Y_1$, special consideration needs to be taken. As imputation has been applied up to time $t_{j-1}$, we assume all data up to time $t_{j-1}$ are not missing and not censored. First, we fit an exponential survival model for the relative time of $\min(t_j-t_{j-1},T_1-t_{j-1})$ and censoring status $I\{T_1>t_j \mbox{ or } (t_{j-1} < T_1 \le t_j \mbox{ and } C_1 = 1)\}$ with independent variables up to time $t_{j-1}$, by only including subjects with $T_1>t_{j-1}$ in which the hazard function is modeled as
    \begin{equation} \label{eq:exp_surv}
    \lambda(t) = \exp\left( (\ddot W^{(j-1)})' \theta_1^{(j)} \right),
    \end{equation}
    where $\ddot W^{(j)} = \left(1, X', (\ddot Z^{(1*)})', (\ddot Z^{(2*)})', \ldots, (\ddot Z^{(j*)})'\right)'$ 
    and $\ddot Z^{(j*)} = \left(Z_2^{(j*)}, Z_3^{(j)}, \ldots, Z_K^{(j)}\right)'$.
   
    Note: for subjects with $T_1>t_{j-1}$, $Z_1^{(1*)}, Z_1^{(2*)}, \ldots, Z_1^{((j-1)*)}$ are all equal to 0, therefore, these variables are not included in the exponential regression model (\ref{eq:exp_surv}). Let $\hat\theta_1^{(j)}$ be the estimator for $\theta_1^{(j)}$, and $V_1^{(j)} = \widehat{\text{Var}}(\hat\theta_1^{(j)})$ be the corresponding variance estimator. It follows that the estimator for the hazard rate is given by $\hat\lambda_1^{(j)} = \exp\left\{(\ddot W^{(j-1)})' \hat \theta_1^{(j)}\right\}$. 
    Then, generate a random variable $\tilde\theta_1^{(j)}\sim \mathcal{N} \left(\hat\theta_1^{(j)},V_1^{(j)}\right)$. For subjects with censoring occurring before $t_j$, $z_1^{(j)}$ (time to event related to $t_{j-1}$) is generated from an exponential distribution with the adjusted hazard rate  
    \begin{equation} \label{eq:bias_adj}
    \ddot \lambda_1^{(j)} = \tilde \lambda_1^{(j)} \exp\left(- (\ddot W^{(j-1)})' V_1^{(j)} (\ddot W^{(j-1)}) /2\right),
    \end{equation}
    where $\tilde\lambda_1^{(j)} = \exp\left\{(\ddot W^{(j-1)})' \tilde \theta_1^{(j)}\right\}$. 
    In Equation (\ref{eq:bias_adj}), a bias correction is applied because $\tilde \lambda_1^{(j)}$ is biased for $\hat \lambda_1^{(j)}$ even though $\log(\tilde \lambda_1^{(j)})$ is not biased for $\log(\hat \lambda_1^{(j)})$ due to the non-linearity of the exponential function. Based on the moment generating function of multivariate normal distribution, we have
    \begin{align*}
        E\left[\tilde \lambda_1^{(j)}|\hat \lambda_1^{(j)}, \ddot W^{(j-1)},  V_1^{(j)} \right] &= E \left[\exp\left\{ (\ddot W^{(j-1)})' \tilde \theta_1^{(j)} \right\} \right] \\
                &= \exp \left\{ (\ddot W^{(j-1)})'\hat \theta_1^{(j)} + {(\ddot W^{(j-1)})}' V_1^{(j)} (\ddot W^{(j-1)} ) /2 \right\} \\
        &= \hat \lambda_1^{(j)} \exp\left((\ddot W^{(j-1)})' V_1^{(j)} (\ddot W^{(j-1)} ) /2\right).
    \end{align*}
    Therefore,
    $ \ddot \lambda_1^{(j)} = \tilde \lambda_1^{(j)} \exp\left(- (\ddot W^{(j-1)})' V_1^{(j)} (\ddot W^{(j-1)}) /2\right)$ is the unbiased estimator for $\hat \lambda_1^{(j)}$. Let $t_c^*=\max(T_c, t_{j-1})$. If $z_1^{(j)} < t_j-t_c^*$, the imputed event occurs in the $j$-th interval and the imputed event time is $t_c^*+z_1^{(j)}$; otherwise, the imputation procedure in this step will be repeated for the next time interval. 

    \item
    For the recurrent event variable $Y_2$, a similar approach based on a piecewise rate function can be used. Assuming the event rate is constant in $j$-th interval, the number of events $Z_2^{(j,*)}$ in the $j$-th time interval can be modeled through a negative binomial regression with mean
    \[
    \log\left\{E\left[Z_2^{(j,*)}|W^{(j-1)}\right]\right\} = \left(W^{(j-1)}\right)'\theta_2^{(j)} + \log\left\{\min(t_j, T_c)-t_{j-1}\right\},
    \]
    where $W^{(j-1)} = \left(1, X', Z^{(1*)'}, Z^{(2*)'}, \ldots, Z^{(j-1,*)'}\right)'$ and $\log\left\{\min(t_j, T_c)-t_{j-1}\right\}$ is the offset parameter. Generate a random variable $\tilde \theta_2^{(j)} \sim \mathcal{N} \left(\hat \theta_2^{(j)}, \widehat{\text{Var}}(\hat \theta_2^{(j)})\right)$, where $\hat \theta_2^{(j)}$ is the estimator for $\theta_2^{(j)}$ and $\widehat{\text{Var}}(\hat \theta_2^{(j)})$ is the variance estimator, by including data for patients with $T_c>t_{j-1}$. A random number $z_2^{(j)}$ is generated from the exponential distribution with rate parameter 
    \begin{equation} \label{eq:bias_adj_nu1}
    \ddot \lambda_2^{(j)} = \frac{1}{\Delta t_j} \exp\left\{\left(W^{(j-1)}\right)' \tilde \theta_2^{(j)} - (W^{(j-1)})' V_2^{(j)} (W^{(j-1)}) /2\right\}.
    \end{equation}
  Let $t_c^*=\max(T_c, t_{j-1})$. If $z_2^{(j)} < t_j-t_c^*$, the imputed event occurs in the $j$-th interval and the imputed event time is $t_{c}^*+z_2^{(j)}$. Then, set the new ``censoring time" $t_c^*$ to be $t_c^*+z_2^{(j)}$ and repeat the above process until the newly generated random number for relative event time is equal to or greater than $t_j-t_c^*$.  
\end{enumerate}

In the above imputation process, we have assumed $Y_1$ is not the terminal event. If $Y_1$ is a terminal event, we can do a post-imputation process by removing all data imputed for time points after the terminal event.

{A few approaches for the multi-level ordinal and categorical variables. One can impute the ordinal variable as if it were a continuous variable and then categorize the imputed value. Then, the problem of imputing the ordinal variable is converted to the problem of imputing a continuous variable. More details and techniques regarding using Caussian-based imputation routines for ordinal variables can be found in \cite{yucel2011gaussian}. The ordinal data can also be modeled by a proportional odds model, which is similar to the logistic regression used for the binary variable. For a multi-level categorical variable, the multinomial logit model can be used \citep{van2018flexible}.}

{The imputation models for the continuous and binary variables are based on well-established methods.  For the TTE variable, a new bias correction method in Equation (\ref{eq:bias_adj}) is introduced compared to the method in \cite{murad2020imputing}. For the TTRE variable, a bias correction method in Equation (\ref{eq:bias_adj_nu1}) is introduced and the time-dependent covariates is included to allow modeling subject-level event rates compared to the method in \cite{tang2018algorithms}.  
}

\section{Simulation} \label{sec:simulation}

\subsection{Data generation}
We consider 2 treatment groups and 4 variables. Let $A$ denote the treatment group ($A=0$ for the control group and $A=1$ for the experimental treatment group). Let $Y_1$ denote a TTE variable, $Y_2$ for a TTRE variable, $Y_3$ for a continuous variable, and $Y_4$ is a binary variable. Assume we have longitudinal measurements for $Y_3$ and $Y_4$ at time $0=t_0<t_1<\ldots<t_J$. The data-generation models are described as follows:
\begin{enumerate}
    \item $A \sim \text{Bernoulli}(0.5)$
    \item A baseline covariate $X \sim \mathcal{N}(\mu_{X}, \sigma_{X}^2)$
    \item For variable $Y_3$ and $Y_4$, we first generate $Y_{kj}^*$ from an integrated two-component prediction model \citep{fu2010bayesian, qu2019can}: 
\begin{equation} \label{eq:ITP}
Y_{kj}^* = m_{k}^*(t_j) + \epsilon_{kj}, \quad k=3,4; j=0,1, \ldots, J,    
\end{equation}
where $$m_{k}^*(t) = \beta_{k0} + \beta_{k1} X + ( \beta_{k2} A  + s_{k}) (1-e^{-\kappa_k t}),$$ $\kappa_k>0$ controls the rate of change over time, $\beta_k=(\beta_{k0}, \beta_{k1}, \beta_{k2})'$ is the vector of fixed effects, $s_{k} \sim \mathcal{N}(0, \sigma_{ks}^2)$ is the random effect to model the between-subject variability, $\epsilon_{kj} \sim \mathcal{N}(0 ,\sigma_{k\epsilon}^2)$ is the residual error, and the correlation between $s_3$ and $s_4$ is $\rho$. Then, we apply transformations $g_3$ and $g_4$ to get the $Y_3$ and $Y_4$ such that
\[
Y_{kj} = g_k(Y_{kj}^*), \quad k=3,4; j=0,1,\ldots, J.
\]
We assume $Y_{3j}$ follows a Gaussian distribution, so the transformation function $g_3$ is an identity function $g_3(x)=x$. The step function
\[
g_4(x) = \left\{ \begin{array}{cl} 0 & \mbox{ if } x < 0 \\ 1 &  \mbox{ if } x \ge 0 \end{array} \right. 
\] is used to define the binary variable $Y_4$.
\item The hazard function for $Y_1$ is given by $$\lambda_1(t) = \lambda_{10} \exp\{\alpha_{11}A + \alpha_{12}X + \alpha_{13} m_{3}^*(t) + \alpha_{14}m_{4}^*(t)\},$$ where $\alpha_{13}$ and $\alpha_{14}$ generally have the opposite sign compared to $\beta_{31}$ and $\beta_{41}$, respectively. Then, we can apply the methods in Austin (2012) to simulate $T_1^*$, the time to event without censoring. The cumulative hazard function can be expressed as
$$\Lambda_1(t) = \int_0^t \lambda_{10} \exp\{\alpha_{11}A + \alpha_{12}X + \alpha_{13} m_{3}^*(v) + \alpha_{14}m_{4}^*(v)\} \} \text{d}v $$
Then, the survival function is given by
\begin{eqnarray}
S(t) &=& \exp\left[ -\int_0^t \lambda_{10} \exp\{\alpha_{11} A + \alpha_{12} X + \alpha_{13} m_j^*(v) + \alpha_{14}m_{4}^*(v)\} \text{d}v \right] \nonumber \\
&=& \exp\bigg\{ - \lambda_{10}  \exp\{(\alpha_{13}\beta_{30} + \alpha_{14}\beta_{40}) + \alpha_{11} A + (\alpha_{12} + \alpha_{13}\beta_{31} + \alpha_{14}\beta_{41})X \} 
\nonumber \\ && \quad\quad
\int_{0}^{t} \exp\{\alpha_{13} (\beta_{32} A + s_{3}) (1-e^{-\kappa_{3}v}) + \alpha_{14} (\beta_{42} A + s_{4}) (1-e^{-\kappa_{4}v})\}\text{d}v\bigg\}. \nonumber
\end{eqnarray}
To generate the random number for the survival function $S(\cdot)$, we can generate a random number $u$ from uniform distribution in $[0,1]$,  then generate survival time from the inverse function 
$T_1 = S^{-1}(u)$. In general, the inverse function $S^{-1}$ needs to be calculated by a numeric method.
\item The rate function for $Y_2$ is given by 
$$\lambda_2(t) = \lambda_{20} \exp\{\alpha_{21}A + \alpha_{22}X + \alpha_{23} m_{3}^*(t) + \alpha_{24}m_{4}^*(t)\}. $$
The time to recurrent event $Y_2$ can be generated in a similar fashion as $Y_1$, except repeatedly generating time from the previous to next event until reaching the last time point $t_J$. For every recurrent event, the cumulative hazard function accumulates from the start of the previous event.
\end{enumerate}


We let the study treatment period be 12 months. To generate baseline covariate $X$, we let $\mu_x=0$ and $\sigma_x=1$. For the longitudinal data, $Y_3$ and $Y_4$ are assumed to be measured at baseline, 3, 6, 9, and 12 months. For the parameters in the longitudinal model (\ref{eq:ITP}), let $\beta_3 = (0, 1, -0.5)'$,  $\beta_4 = (0, 1, -0.5)'$, $k_3 = 0.5$, $k_4 = 0.15$, $\sigma^2_{3s} =0.1$, $\sigma^2_{4s} = 0.1$, $\sigma^2_{3\epsilon} = \sigma^2_{4\epsilon} = 0.4$, and $\rho = 0.5$. For the clinical outcomes, let $\lambda_{10} = 0.08$, $\lambda_{20} = 0.13$, and $\alpha_1 = \alpha_2 = (-0.7, 0.5, 0.5, -0.5)'$. 

We consider 2 censoring mechanisms: independent censoring and dependent censoring. In both mechanisms, we assume the censoring occurs only at 3, 6, 9, and 12 months. 
The independent censoring indicators are generated from a Bernoulli distribution with the probabilities of 0.15, 0.2, 0.25, and 0.4 at 3, 6, 9, and 12 months, respectively. For dependent censoring, the probability of being censored $\pi^{(j)}$ is generated from a logistic model:
\[
\mbox{logit}(\pi^{(j)}) = -1 + 0.8 Y_{3j} - 0.5 Y_{4j}. 
\]
In both censoring scenarios, approximately 60\% of patients are censored before 12 months and the remaining patients are censored at 12 months. 

\subsection{Simulation results}
We compare the mean survival curve based on the Kaplan-Meier estimator for the TTE outcome ($Y_1$), mean cumulative function (MCF) based on the Nelson-Aalen estimator \citep{nelson2003recurrent} for the TTRE outcome ($Y_2$), the mean for the continuous outcome $Y_3$, and the mean proportion of response for the binary outcome ($Y_4$) based on the proposed imputation method and the naive estimator (ignoring missing values). In addition, we also present the estimates with complete data (without censoring), which serves as a gold standard. All results were based on 2000 simulations with a sample size of 500. Patients were randomly assigned to treatment group $A=0$ or 1 with probability of 0.5. For each simulated data, 50 imputed data sets were generated. In this simulation, we include treatment group indicator as a covariate. When the sample size is larger, we can make the imputation model more flexible by performing the imputation separately for each treatment group.

\begin{figure}[H]
\centering
\includegraphics[scale=0.6]{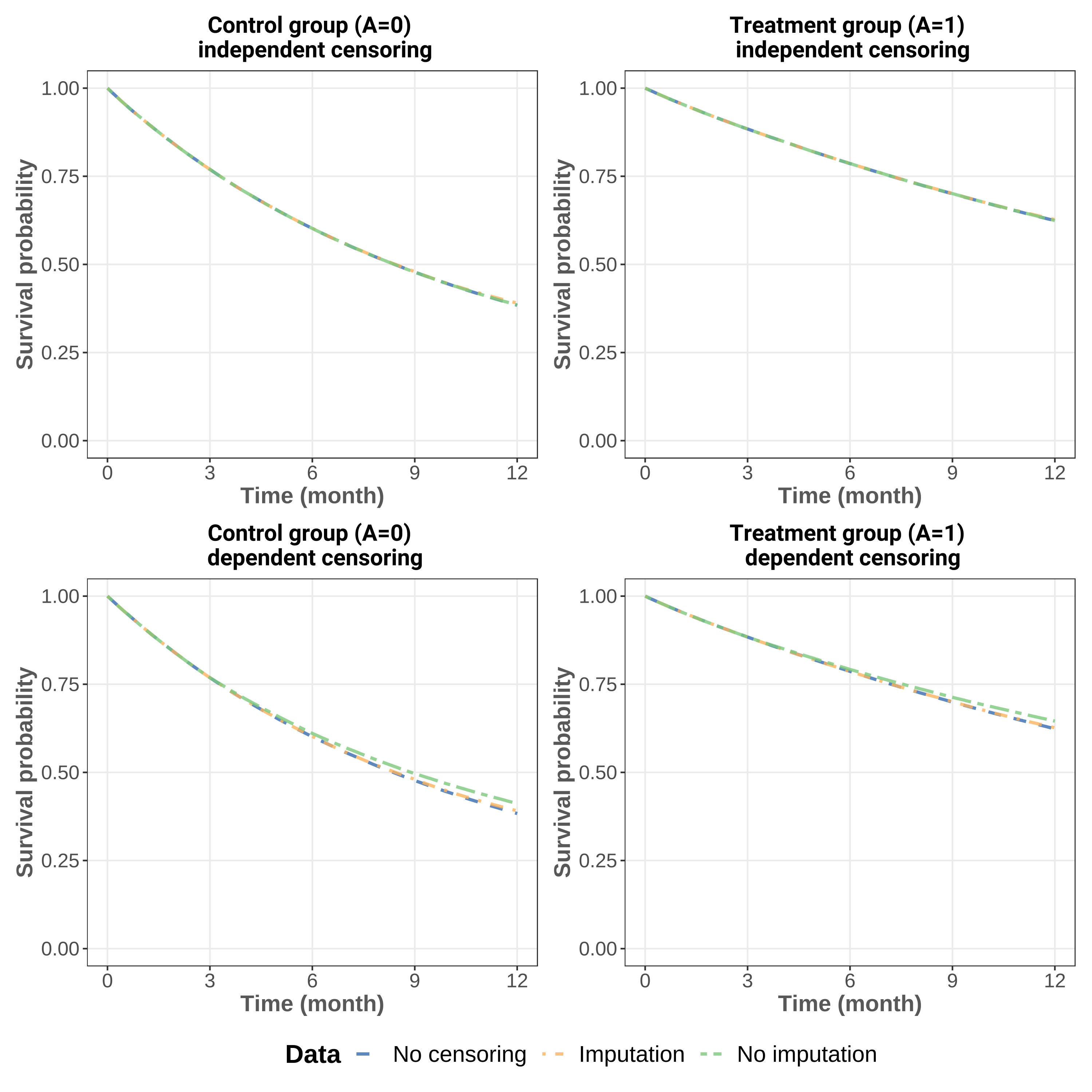}
\caption{ \small
Mean survival curves based on the Kaplan-Meier estimator for data with independent and dependent censoring. 
}
\label{plot:rel_sf}
\end{figure}

\begin{figure}[H]
\includegraphics[scale=0.6]{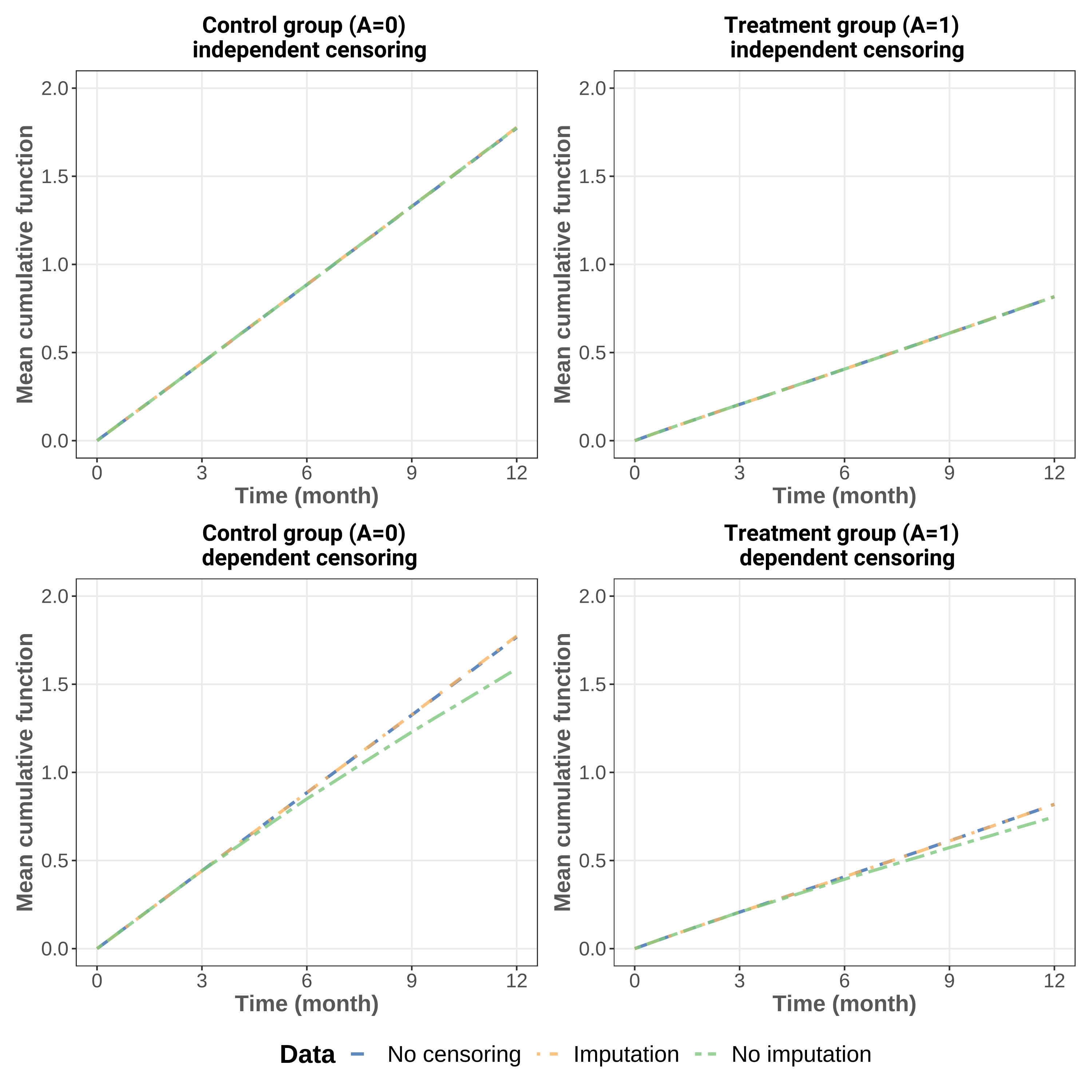}
\caption{ \small
Mean cumulative curves based on the Nelson-Aalen estimator for data with independent and dependent censoring. 
}
\label{plot:rel_mcf}
\end{figure}

\begin{figure}[H]
\includegraphics[scale=0.6]{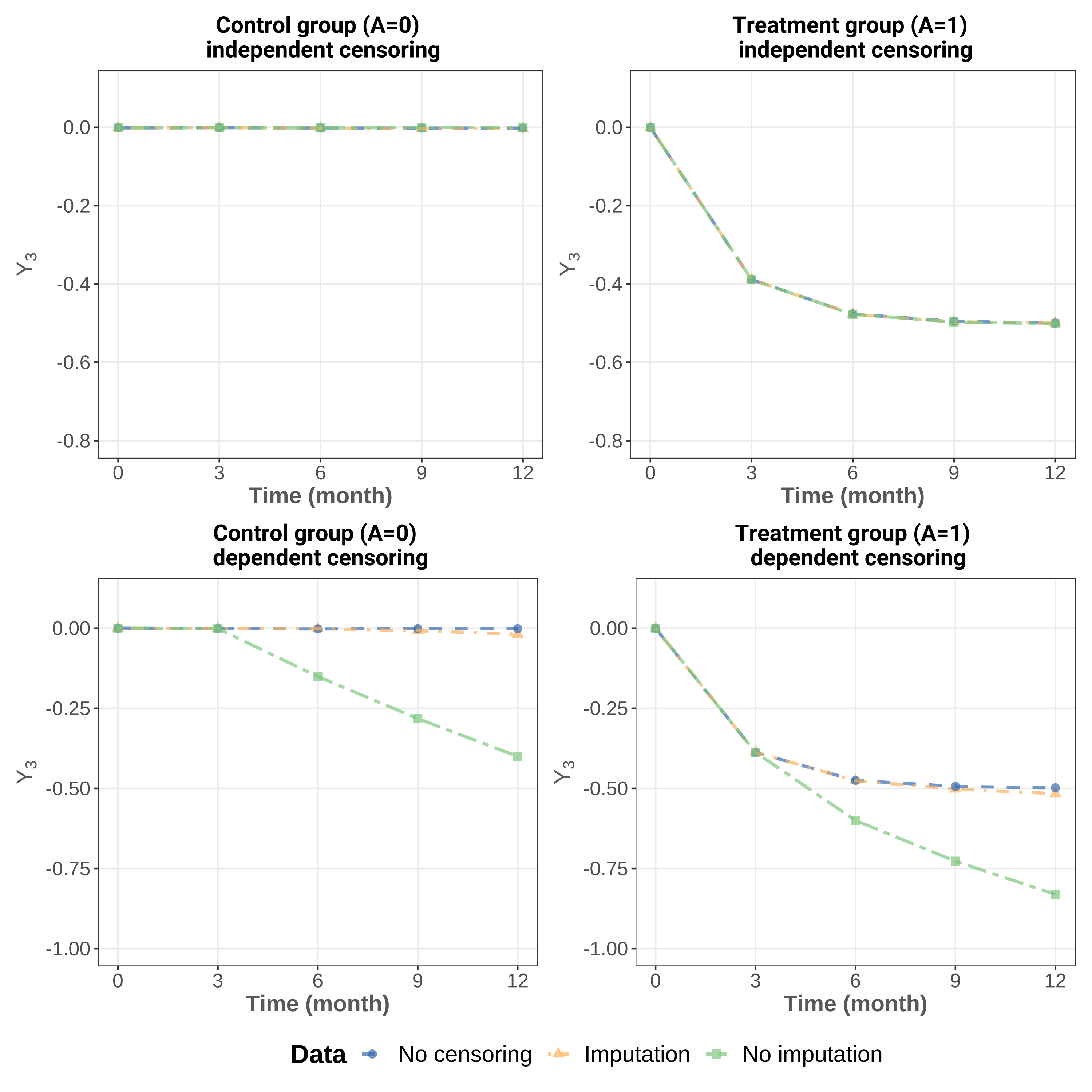}
\caption{ \small
Mean of the continuous variable for data with independent and dependent censoring. 
}
\label{plot:rel_y}
\end{figure}

\begin{figure}[H]
\includegraphics[scale=0.6]{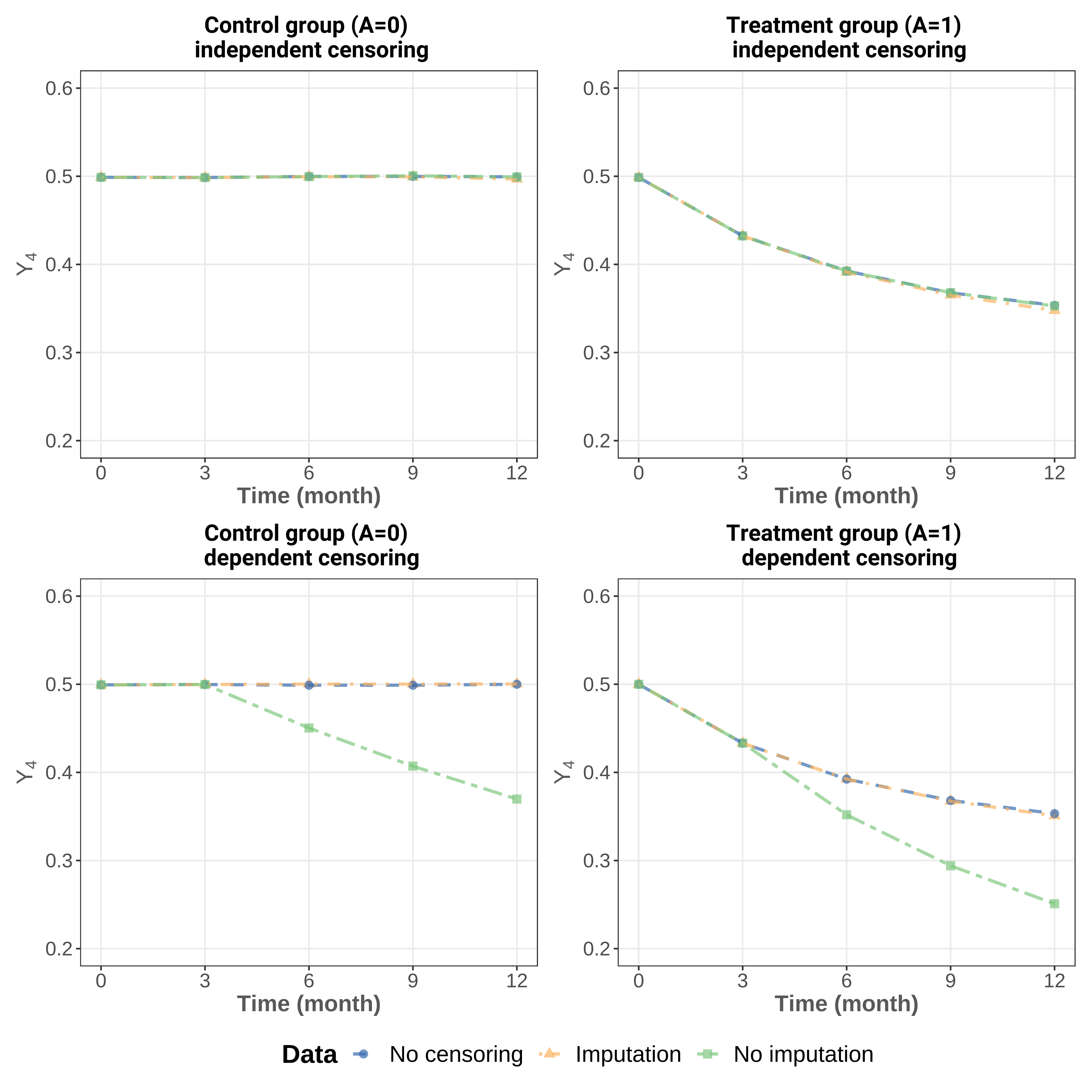}
\caption{ \small
Mean of the binary variable for data with independent and dependent censoring. 
}
\label{plot:rel_b}
\end{figure}

Figures \ref{plot:rel_sf}-\ref{plot:rel_b} show the simulation results averaged across the 2000 simulation samples. Blue points and curves are the estimates based on data without censoring, green points and curves for estimates based on data with censoring but without imputation, and brown points/curves for the estimates with the imputation of the censored data. 

Figures \ref{plot:rel_sf} shows the mean survival curve for each treatment for the cases of independent and dependent censoring. For independent censoring, all 3 estimates were almost perfectly overlapped. For dependent censoring, the mean survival curve was clearly biased when using the Kaplan-Meier estimator directly without imputation. The mean survival curve estimated after imputation showed very little bias. 

Figure \ref{plot:rel_mcf} shows the average MCF over the 2000 estimated MCF curves. Again, for independent censoring, the MCF estimated using the Nelson-Aalen estimator without imputation estimates the MCF well. For dependent censoring, the Nelson-Aalen estimator was apparently biased. The estimated MCF after imputation was almost perfectly overlapped with the calculated MCF without censoring, indicating the imputation performs well in estimating the MCF. 

Figures \ref{plot:rel_y} and \ref{plot:rel_b} show the mean responses over time for the continuous and binary variables, respectively. Similarly, the mean responses based on imputed data had little bias for both independent and dependent censoring.

The inferences for estimates using multiple imputed data can be drawn based on the variance estimation using Rubin's Rule \citep{rubin1987multiple,barnard1999miscellanea}. For any estimand $\theta$, let $\widehat{\theta}_i$ be the estimator for $\theta$ and $\widehat{V}_i$ be the corresponding variance estimator from $i$-th imputed data ($i=1,2,\ldots,m$). Then, the estimator combining estimates from multiple imputation $\bar{\theta} = \frac{1}{m}\sum_{i=1}^{m} \theta_i$ and corresponding variance estimator using Rubin's Rule is given by
$$V_{\text{pooled}} = V_{\text{within}} + \left(1 + \frac{1}{m}\right) V_{\text{between}},$$
where $V_{\text{within}} = \frac{1}{m}\sum_{i=1}^{m}\widehat{V}_i$ and $V_{\text{between}} = \frac{1}{m-1}\sum_{i=1}^{m}(\widehat{\theta}_i - \bar{\theta})^2$. 

We evaluated the performance of the estimators for the survival probability at $t=12$, the number of recurrent events during time interval $(0,12]$, the mean response for $Y_3$ at $t=12$, and the probability of $Y_4=1$ at $t=12$. For each simulated (or imputed) data, the parameters were estimated as follows:
\begin{itemize}
    \item For survival probability at $t=12$,  the Kaplan-Meier estimator was used to calculate $\widehat{\theta}_i$ and $\widehat{V}_i$ was calculated using the Greenwood formula.
    \item  For the number of recurrent events from (0, 12], we used Nelson-Aalen type estimator to calculate $\widehat{\theta}_i$ and $\widehat{V}_i$.
    \item  For the mean value of $Y_3$ at 12, $\widehat{\theta}_i$ is the average value of $Y_3$ and $\widehat{V}_i$ is the variance of $Y_3$ divided by $n$.
    \item  For the probability of $Y_4 = 1$ at 12, $\widehat{\theta}_i$ is the average value of $Y_3$ and $\widehat{V}_i = \frac{\widehat{\theta}_i(1-\widehat{\theta}_i)}{n}$.
\end{itemize}

Table \ref{table:independent} provides the mean and standard deviation of the estimates, the mean estimated standard errors and the coverage probability of the 95\% confidence interval based on the data after imputation for the case of independent censoring. The mean and standard deviation based on the data without censoring, which serves the gold standard, are also reported. For all parameters, the mean estimates based on multiple imputation were similar to those without censoring. For multiple imputation, the mean standard errors were similar to the standard deviations of the parameter estimates, and the 95\% confidence intervals had approximately 95\% coverage probability. For all parameters, the standard deviations of the estimates based on multiple imputation were larger than those based on data without censoring, which makes sense as censoring results in losing information.

 
Table \ref{table:dependent} shows the simulation results for the case of dependent censoring. Similarly, the estimates seemed to have little bias,  the estimated standard errors were close to the standard deviation of the estimates, and the 95\% confidence intervals had the appropriate coverage probability. 

\begin{figure}[H]
\includegraphics[scale=0.6]{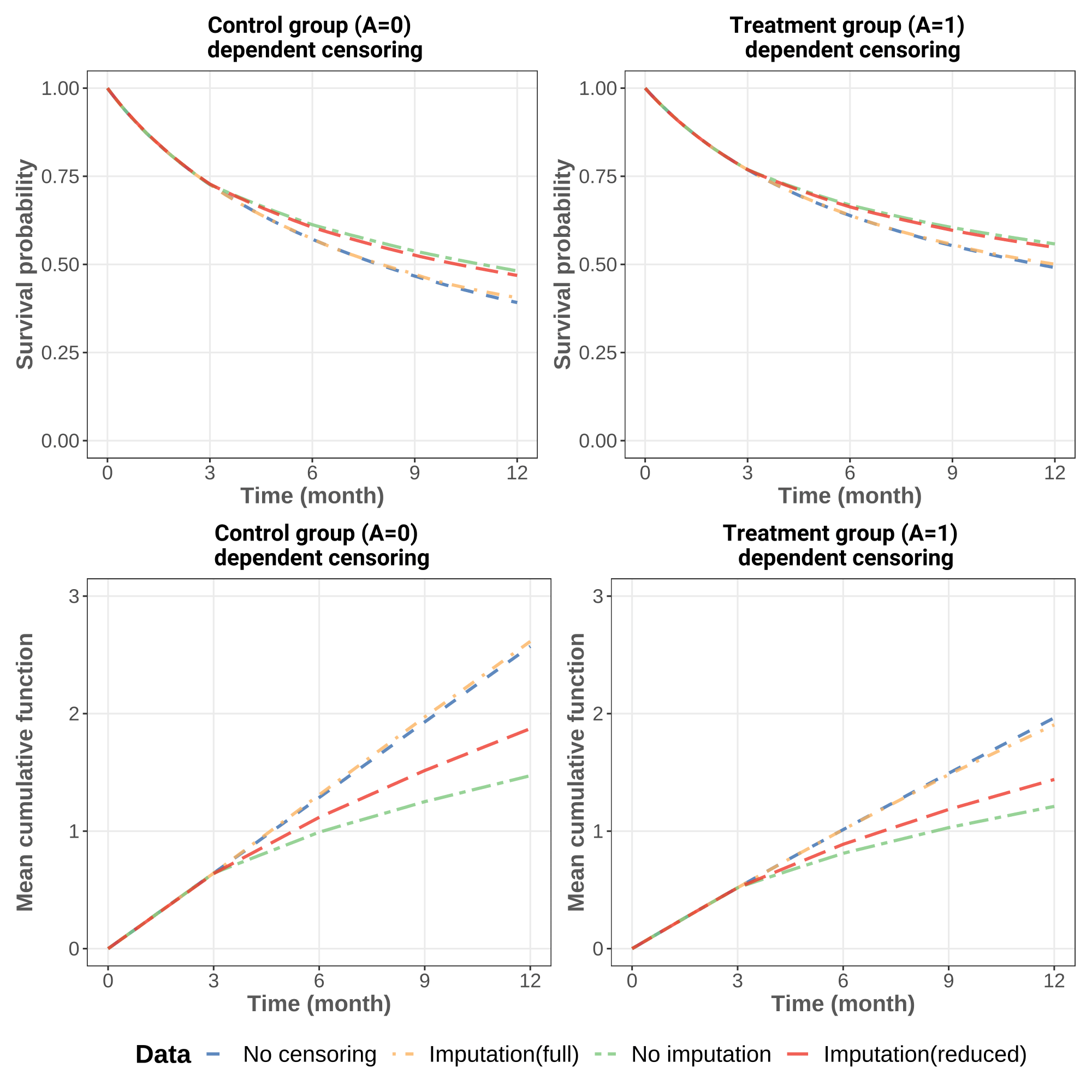}
\caption{ \small
Mean survival curves and mean cumulative curves for data with dependent censoring (comparing imputation models with full and reduced dependencies)}
\label{plot:clinical}
\end{figure}

\begin{figure}[H]
\includegraphics[scale=0.6]{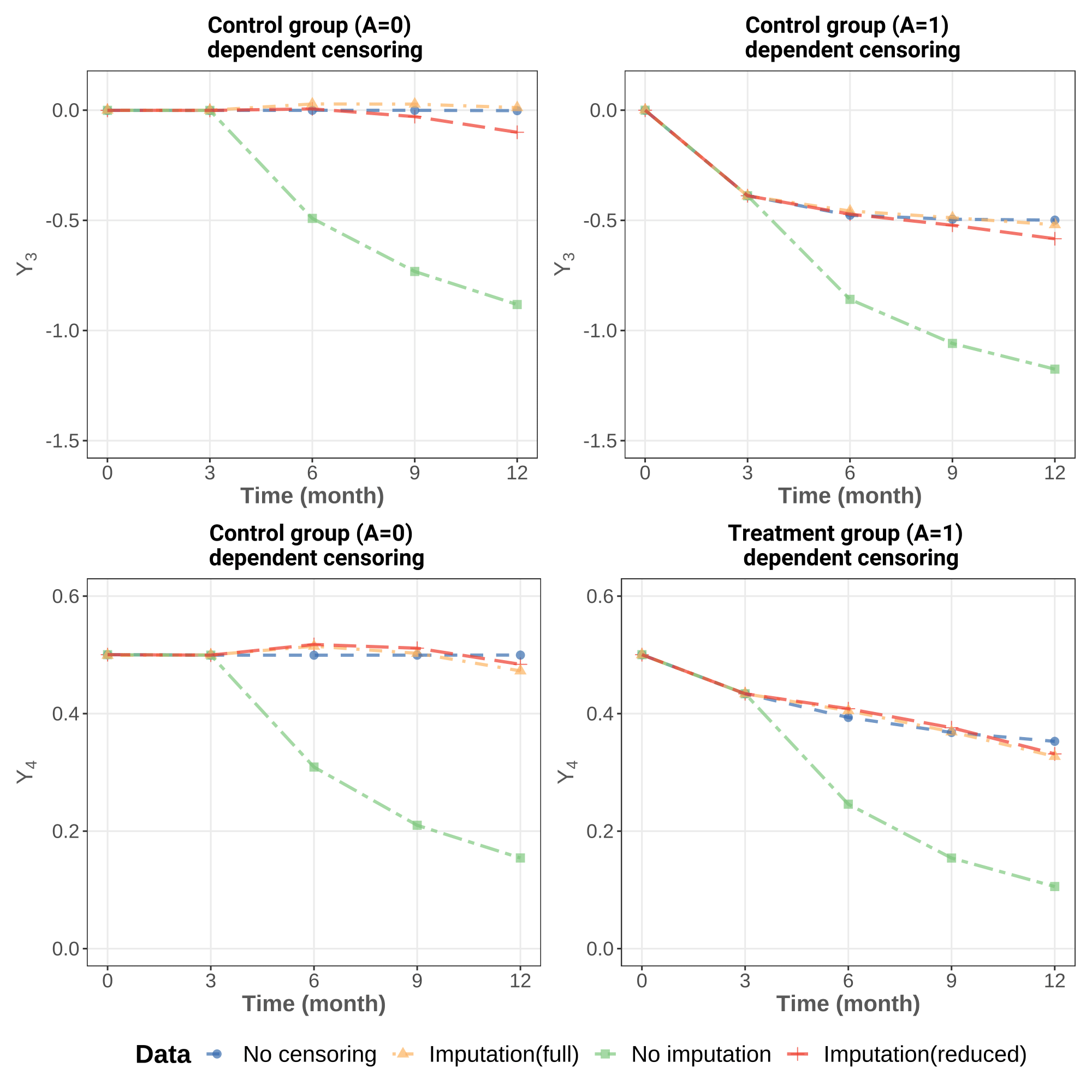}
\caption{ \small
Mean of the continuous and binary variables for data with dependent censoring (comparing imputation models with full and reduced dependencies)}
\label{plot:longitudinal}
\end{figure}

We conducted additional simulations to compare the performance of the proposed imputation algorithm based on full model described in Section 2 and the reduced model without considering the dependencies between different types of variables. We followed the set-up in Section 3.1 with modifications on the values of the parameters in the clinical outcomes and censoring model to allow a stronter between-variable dependency. Let $\alpha_1 = \alpha_2 = (-0.2, 0.0, -0.5, 1.5)'$.  For dependent censoring, the probability of being censored $\pi^{(j)}$ is generated from a logistic model $\mbox{logit}(\pi^{(j)}) = -1 + 1.8 Y_{3j} + 0.5 Y_{4j}$. The imputation based on full model contains both clinical outcomes and longitudinal outcomes. For the imputation based on reduced model, we excluded longitudinal outcomes when imputing clinical outcomes and excluded clinical outcomes when imputing longitudinal outcomes. The performance of the imputation based on full and reduced models are summarized in Figure \ref{plot:clinical} and Figure \ref{plot:longitudinal}. When imputing the clinical outcomes, the full model produced unbiased results, while the reduced model produced considerable bias. For the longitudinal outcomes, the difference between the full model and reduced model is minor, which is reasonable as the longitudinal data are generated independently from the clinical outcome.

\section{Application} \label{sec:application}
We applied the proposed imputation method to data from the IMAGINE-3 Study (ClinicalTrials.gov Identifier: NCT01454284), a parallel, double-blind, 52-week, multi-center, phase 3 study in patients with type 1 diabetes mellitus. Eligible patients were randomly assigned to insulin peglispro or insulin glargine (two long-acting insulin formulations), with the addition of  short-acting insulin used to control the postprandial glucose level or to correct high glucose at any time. In this trial, 1112 patients were randomized and took at least one dose of randomized study medication (663 in insulin peglispro and 449 in insulin glargine). Insulin doses were adjusted weekly in the first 12 weeks of treatment and then adjusted according to investigators' judgement thereafter. The study showed insulin peglispro treatment resulted in significantly better glycemic control with lower hemoglobin A1c (HbA1c) and fasting serum glucose (FSG) compared to insulin glargine \citep{bergenstal2016randomized}. However, insulin peglispro was associated with increased triglycerides and alanine transaminase (ALT) compared to insulin glargine. Additionally, insulin peglispro was also associated with increased total hypoglycemic events, especially at the beginning of the study.

In this section, we applied the proposed imputation method to impute the continuous variables of HbA1c, FSG, ALT, and triglycerides at Week 4, 12, 26 and 52, and to impute the time to recurrent events for documented symptomatic hypoglycemia defined by glucose value less than 70 mg/dL and accompanied with symptoms. During the study, 18\% of patients treated with insulin glargine and 23\% of patients treated with insulin peglispro discontinued the study medication.  We used a hypothetical strategy to handle treatment discontinuations, i.e., censoring data after treatment discontinuation. As these variables could be correlated, 
it makes sense to consider all these variables together when imputing the data. The missing data were imputed 100 times and the mean estimates were the average of estimates from 100 imputed samples, and the standard errors were calculated using Rubin's rule of combining within- and between-imputation variabilities. Since our purpose is to illustrate the imputation method, we performed the imputation by treatment group and compared the estimates with traditional methods within each treatment group (without focusing on the between-treatment comparison). 

\begin{figure}[htbp]
\centering
\includegraphics[scale=0.36]{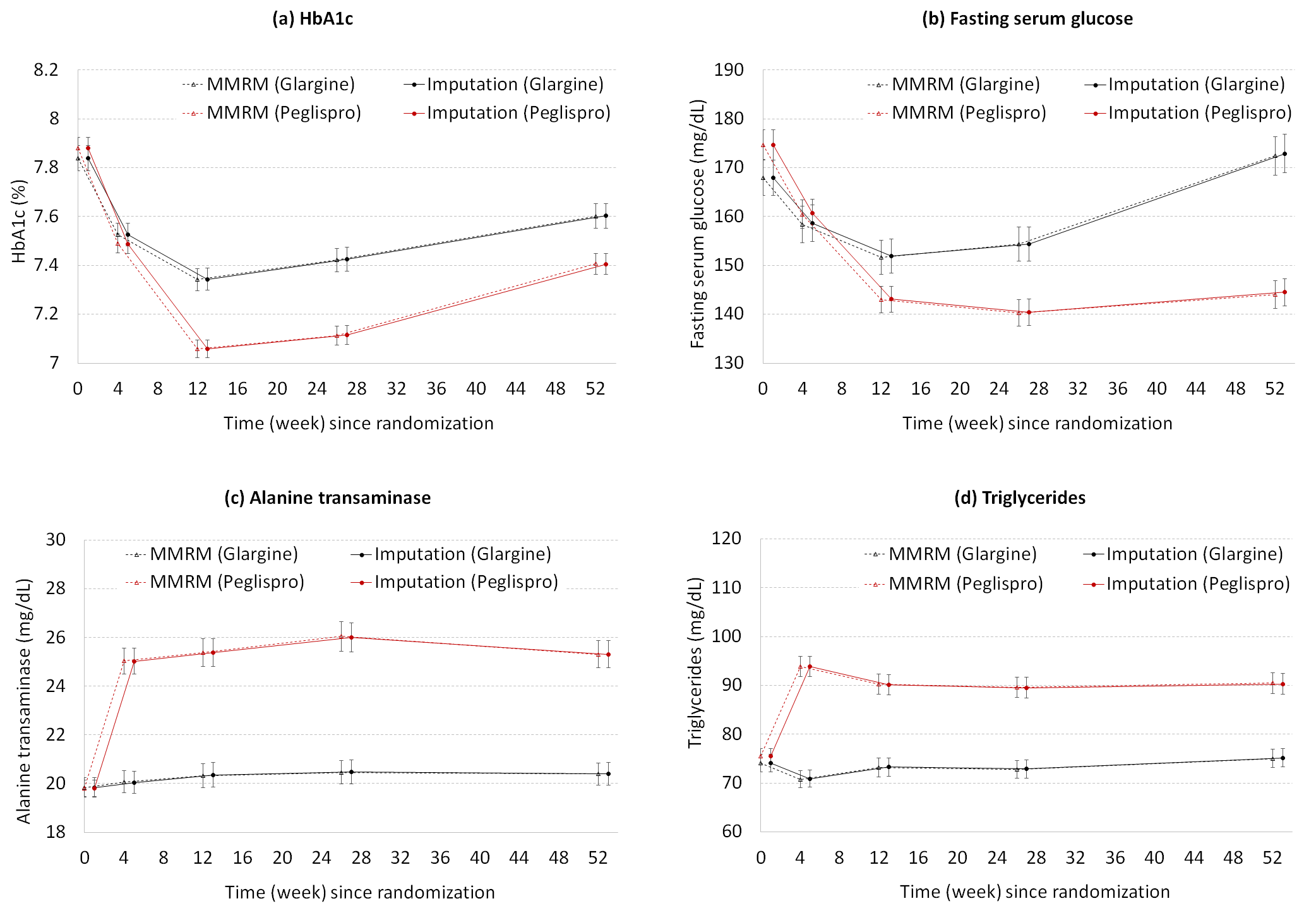}
\caption{ 
Means and standard errors for HbA1c, fasting serum glucose, alanine transminase, and triglycerides based on MMRM and multiple imputation. 
}
\label{plot:labs}
\end{figure}

Figure \ref{plot:labs} shows the mean and standard error for the continuous variables, compared to the mean and standard error estimated from the mixed model for repeated measures (MMRM) with factors of time point within each treatment group. The results from the two methods were very similar for all variables across all time points. Since the MMRM for each variable assumes the probability of missingness at most only depends on the previous observed values for this variable (equivalent to the multiple imputation using only the longitudinal data for this variable), this means that the other variables did not provide much additional information in imputing the missing values for this variable. 

\begin{figure}[htbp]
\centering
\includegraphics[scale=0.35]{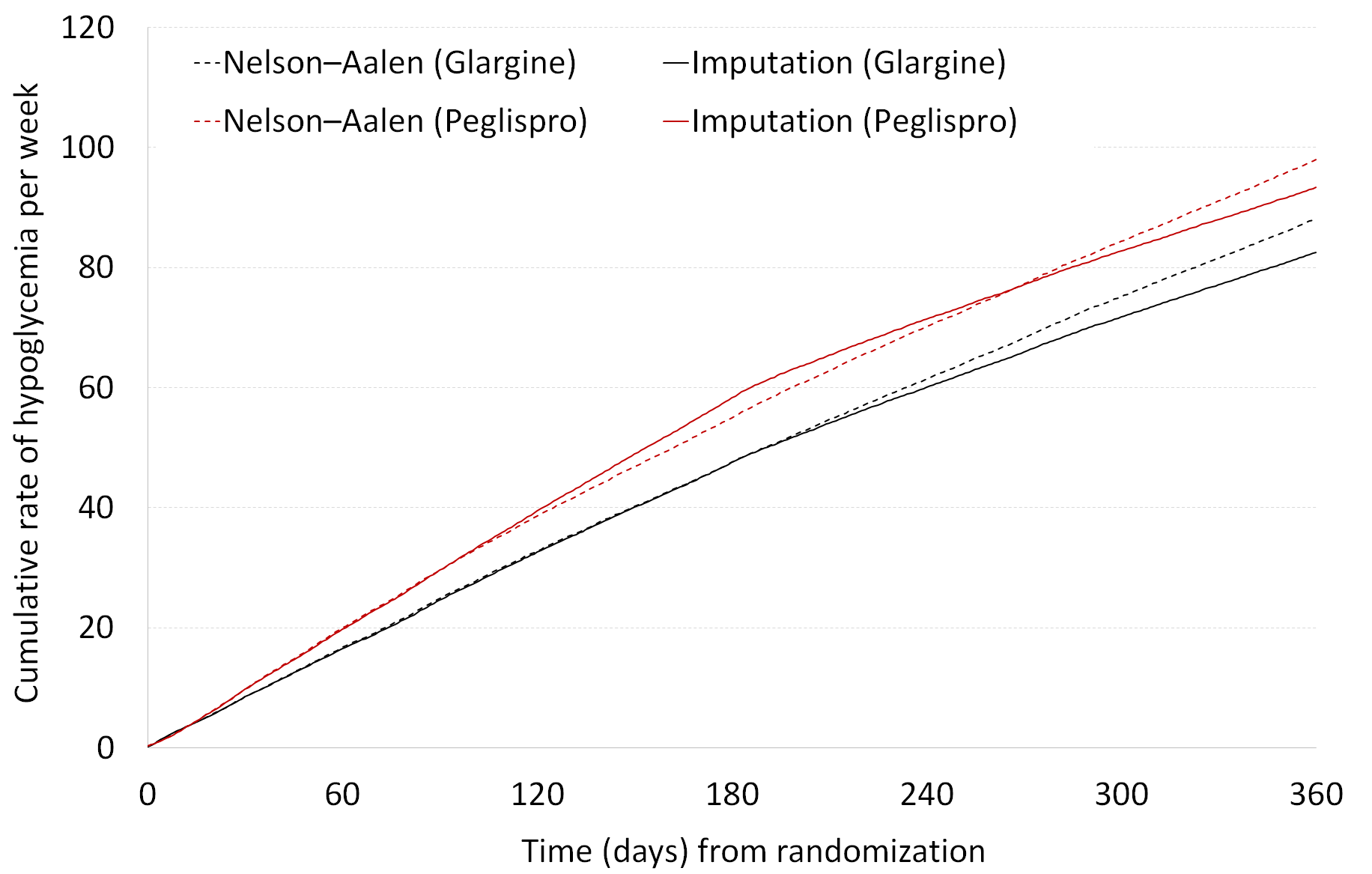}
\caption{ 
Mean cumulative hypoglycemia event rate (per 7 days). 
}
\label{plot:hypo}
\end{figure}

Figure \ref{plot:hypo} shows the estimated mean cumulative hypoglycemia rate function based on the Nelson-Aalen estimator and multiple imputation, respectively. For the insulin glargine group, the cumulative hypoglycemia event rate based on multiple imputation was at first similar to the event rate based on the Nelson-Aalen estimator, but became lower after 26 weeks. Further investigation suggested patients who discontinue the treatment after 26 weeks seemed to have lower hypoglycemia rate compared to those who did not discontinue treatment after 26 weeks. For insulin peglispro group, the cumulative hypoglycemia event rate based on multiple imputation was initially higher but lower at the end of the study, compared to the event rate based on Nelson-Aalen estimator. With further investigation, we found that patients who discontinued the study earlier tended to have higher hypoglycemia events and patients who discontinued the study later tended to have lower hypoglycemia, compared to those who adhered to the treatment throughout the study. 

\section{Summary and Discussion} \label{sec:summary}
Missing data are common in clinical trials. The multiple imputation is an important approach to handling missing values. Most imputation packages support the imputation for longitudinal continuous and categorical outcomes, and methods for imputing TTE or TTRE outcomes are not well studied. We proposed a method to impute missing values for a mix of variable types, including TTE, TTRE, continuous, and binary outcomes using an approximate FCS method. {The proposed method may provide an improvement compared to methods imputing each type of variables independently.} 
For imputation of TTE and TTRE outcomes, we assumed a piecewise hazard (or event) rate and modeled the hazard (or event) rate only through the longitudinal outcomes at a few time points. In the simulation studies, although the TTE and TTRE outcomes were generated with the hazard (or event) rate depending on the longitudinal outcome that changes over time, the imputation based on the approximate FCS performed well. Note that even in the simulation study, the conditional distribution used in the imputation is only an approximation as the parametric form of the true conditional distribution is difficult to express based on the data-generation model. This is consistent with the findings by \cite{van2007multiple} that despite the lack of rigorous theoretic properties, the imputation based on FCS generally performs remarkably well. 

\cite{murad2020imputing} proposed a similar method for the TTE outcome when censoring depends on a continuous outcome. The method we proposed in this article has three improvements compared to \cite{murad2020imputing}. First, we allow the distributions of the continuous or binary variable to depend on the TTE and TTRE outcomes. This is especially important because when the longitudinal outcome measures are sparse, the occurrence of (non-terminal) events between times of longitudinal outcome measurements will provide useful information for the imputation of the longitudinal outcome. Second, we applied a bias adjustment for the nonlinear exponential function when imputing the TTE and TTRE variables. This bias adjustment effectively reduced the bias for small to moderate sample sizes in our simulation (results without the bias adjustment not shown). Third, we incorporated a method to impute the TTRE outcome. 

If there are many variables and many time intervals, the number of variables conditioned on could be large and the conditional model may fail to converge in the late intervals due to insufficient number of observations. For example, if there are 10 baseline covariates, 10 postbaseline time intervals, and 10 outcome variables, there will be close to 100 variables to be conditioned on at the 10$^{\mathrm{th}}$ time interval. In this case, we may simplify the model and condition on the most important variables or some composite variables. For example, we may consider a new variable indicating whether the event occurs before the current time for a TTE variable, the total number of recurrent events before the current time for a TTRE variable, and the value only at the prior time for a longitudinal outcome. 

Another way to reduce the complexity of the imputation models is to assume a certain causal relationship between variables as in the structural causal model \citep{spirtes2000causation}. For example,  we can assume a one-direction causal effect for the TTE/TTRE variables and the biomarkers (continuous variables) such that poor biomarkers cause earlier time to (recurrent) events. Then, in the imputation models, the biomarkers can be imputed first without involving the TTE and TTRE variables, which can be done using the standard imputation package for continuous variables. After that, the TTE and TTRE variables can be imputed through a model conditional on the biomarkers.  

If there is more than one longitudinal outcome, the scheduled clinical visits to collect these outcomes may differ. For example, vital signs may be measured at each clinical visit, while some laboratory variables may only be measured at certain visits. The FCS procedure for the longitudinal outcomes does not require all longitudinal measurements. For each longitudinal variable, only the value at the scheduled time may be imputed and be conditioned on. 

We assessed the performance of the imputation when censoring is completely random or depends on the observed longitudinal outcomes. In clinical trials, pattern mixture models are widely used to handle special patterns of missingness, especially for missing values with certain intercurrent events \citep{ich2019e9}. Pattern mixture models can easily be implemented using the proposed imputation method if the goal is to impute certain missing values using data from subjects with a similar pattern. Various sensitivity analyses may be performed by introducing some sensitivity parameters. 

To draw inferences based on imputed data, we evaluated the performance of variance estimation using Rubin's rule by combining the within and between imputation variances \citep{rubin1987multiple,barnard1999miscellanea}. The performance was reasonable for the case of independent censoring and dependent censoring. However, when applying this imputation method to some pattern mixture models where the imputation model and data generation model are uncongenial 
\citep{robins2000inference, hughes2016comparison, bartlett2020bootstrap}, bootstrap methods may be used in estimating the variance \citep{bartlett2020bootstrap}. 

There is one major limitation for the proposed imputation method: it can only be applied to monotone missingness. In the real application, sparse non-monotone missingness may occur, mostly due to technical errors (e.g., the blood sample storing condition is compromised). For sparse non-monotone missing values, one of the below few options can be used:
\begin{enumerate}
\item[$\bullet$] Impute the non-monotone missing outcome using only data collected before this time point.
\item[$\bullet$] For continuous or binary outcome, we may use the linear interpretation to impute the non-monotone missing outcome. For a small number of non-monotone missing values, the lack of the uncertainty in imputed values may have little impact on the inference.
\item[$\bullet$] For continuous or binary outcome, we may impute these values using only the longitudinal outcomes without considering the TTE and TTRE outcomes, which can be directly done using existing imputation packages.
\end{enumerate}
When there is a large amount of non-monotone missingness, a Bayesian approach with fully specification of the joint model may be needed.  

We acknowledge the proposed imputation method could be complex in implementation. We are currently in the process of developing an R package to implement the imputation. The goal is to make the interface of the software as user-friendly as possible, like the MICE R-package, so that applied researchers should be able to use the methodology conveniently in practice.

In conclusion, we have provided a flexible FCS-based imputation method for missing data for a mix of variable types. This may have a wide range of applications in analyzing data from clinical trials. 

\section*{Acknowledgement}

\bibliographystyle{apalike}
\bibliography{references}

\newpage

\begin{table}[H] \centering
\caption{Summary of simulation results for independent censoring}
\footnotesize
\begin{tabular}{ccrccccccc}
\hline\hline
& & & \multicolumn{2}{c}{No Censoring} && \multicolumn{4}{c}{MI} \\
\cline{4-5} \cline{7-10} 
Variable & Group && Mean & SD && Mean & SD & mSE & CP \\ 
\hline 
Survival probability at $t=12$ & Control  && 0.384 & 0.030 && 0.390 & 0.038 & 0.038 & 0.941\\ 
                             & Treatment  && 0.624 & 0.030 && 0.627 & 0.038 & 0.038 & 0.947\\
\hline 
\# of recurrent events for $0<t\le 12$ & Control  && 1.772 & 0.101 && 1.776 & 0.145 & 0.152 & 0.953\\ 
                                         & Treatment  && 0.815 & 0.065 && 0.816 & 0.093 & 0.106 & 0.960\\
\hline 
$E(Y_3)$ at 12 months & Control  && -0.003 & 0.070 && -0.005 & 0.078 & 0.082 & 0.951\\ 
                    & Treatment  && -0.500 & 0.071 && -0.500 & 0.080 & 0.081 & 0.956\\
\hline 
$\Pr(Y_4=1)$ at 12 months & Control  && 0.499 & 0.032 && 0.497 & 0.041 & 0.039 & 0.930\\ 
                        & Treatment  && 0.353 & 0.030 && 0.346 & 0.040 & 0.037 & 0.920\\
\hline\hline
\end{tabular}\\
    {\begin{flushleft} Notation and abbreviations: CP, coverage probability of the 95\% confidence interval; MI, multiple imputation; mSE, mean estimated standard errors based on Rubin's rule, SD, standard deviation.
    \end{flushleft} }
\label{table:independent}
\end{table}

\begin{table}[H] \centering
\caption{Summary of simulation results for dependent censoring}
\footnotesize
\begin{tabular}{ccrccccccc}
\hline\hline
& & & \multicolumn{2}{c}{No Censoring} && \multicolumn{4}{c}{MI} \\
\cline{4-5} \cline{7-10} 
Variable & Group && Mean & SD && Mean & SD & mSE & CP \\ 
\hline 
Survival probability at $t=12$ & Control  && 0.382 & 0.031 && 0.390 & 0.039 & 0.038 & 0.945\\ 
                             & Treatment  && 0.623 & 0.031 && 0.627 & 0.039 & 0.037 & 0.926\\
\hline 
\# of recurrent events for $0<t\le 12$ & Control  && 1.769 & 0.099 && 1.776 & 0.136 & 0.151 & 0.954\\ 
                                         & Treatment  && 0.820 & 0.066 && 0.819 & 0.082 & 0.081 & 0.941\\
\hline 
$E(Y_3)$ at 12 months & Control  && 0.001 & 0.069 && -0.016 & 0.077 & 0.081 & 0.956\\ 
                    & Treatment  && -0.496 & 0.070 && -0.514 & 0.077 & 0.077 & 0.954\\
\hline 
$\Pr(Y_4=1)$ at 12 months & Control  && 0.501 & 0.032 && 0.502 & 0.040 & 0.037 & 0.931\\ 
                        & Treatment  && 0.354 & 0.030 && 0.353 & 0.037 & 0.035 & 0.937\\
\hline\hline
\end{tabular}\\
    {\begin{flushleft} Notation and abbreviations: CP, coverage probability of the 95\% confidence interval; MI, multiple imputation; mSE, mean estimated standard errors based on Rubin's rule, SD, standard deviation.
    \end{flushleft} }
\label{table:dependent}
\end{table}

\end{document}